\documentclass[aps,prb,twocolumn,noeprint]{revtex4-1}

\usepackage[pdftex]{graphics}
\usepackage{amssymb,amsmath}
\usepackage{nicefrac}
\usepackage{color}

\begin{document}
\def\bra#1{\left<{#1}\right|}
\def\ket#1{\left|{#1}\right>}
\def\expval#1#2{\bra{#2} {#1} \ket{#2}}
\def\mapright#1{\smash{\mathop{\longrightarrow}\limits^{_{_{\phantom{X}}}{#1}_{_{\phantom{X}}}}}}

\title{On the low magnetic field effect in radical pair reactions}

\author{Alan M.~Lewis}
\affiliation{Department of Chemistry, University of Oxford, Physical and Theoretical Chemistry Laboratory, South Parks Road, Oxford, OX1 3QZ, UK}

\author{Thomas P. Fay}
\affiliation{Department of Chemistry, University of Oxford, Physical and Theoretical Chemistry Laboratory, South Parks Road, Oxford, OX1 3QZ, UK}

\author{David E.~Manolopoulos}
\affiliation{Department of Chemistry, University of Oxford, Physical and Theoretical Chemistry Laboratory, South Parks Road, Oxford, OX1 3QZ, UK}

\author{Christian Kerpal}
\affiliation{Department of Chemistry, University of Oxford, Centre for Advanced Electron Spin Resonance, South Parks Road, Oxford, OX1 3QR, UK}

\author{Sabine Richert}
\affiliation{Department of Chemistry, University of Oxford, Centre for Advanced Electron Spin Resonance, South Parks Road, Oxford, OX1 3QR, UK}

\author{Christiane R.~Timmel}
\affiliation{Department of Chemistry, University of Oxford, Centre for Advanced Electron Spin Resonance, South Parks Road, Oxford, OX1 3QR, UK}

\begin{abstract}
Radical pair recombination reactions are known to be sensitive to the application of both low and high magnetic fields. The application of a weak magnetic field reduces the singlet yield of a singlet-born radical pair, whereas the application of a strong magnetic field increases the singlet yield. The high field effect arises from energy conservation: when the magnetic field is stronger than the sum of the hyperfine fields in the two radicals, ${\rm S}\to {\rm T}_{\pm}$ transitions become energetically forbidden, thereby reducing the number of pathways for singlet to triplet interconversion. The low field effect arises from symmetry breaking:  the application of a weak magnetic field lifts degeneracies among the zero field eigenstates and increases the number of pathways for singlet to triplet interconversion. However, the details of this effect are more subtle, and have not previously been properly explained. Here we present a complete analysis of the low field effect in a radical pair containing a single proton, and in a radical pair in which one of the radicals contains a large number of hyperfine-coupled nuclear spins.  We find that the new transitions that occur when the field is switched on are between ${\rm S}$ and ${\rm T}_0$ in both cases, and not between ${\rm S}$ and ${\rm T}_{\pm}$ as has previously been claimed. We then illustrate this result by using it in conjunction with semiclassical spin dynamics simulations to account for the observation of a biphasic--triphasic--biphasic transition with increasing magnetic field strength in the magnetic field effect on the time-dependent survival probability of a photoexcited carotenoid-porphyrin-fullerene radical pair.

\end{abstract}

\maketitle

\section{Introduction}

It is well known that the application of a weak magnetic field can have a significant effect on the outcome of a radical pair recombination reaction.\cite{Hamilton1988,Batchelor1993,Stass95a,Stass95b,Timmel2004,Maeda2008} But while this low field effect has been reproduced in theoretical calculations,\cite{Brocklehurst1976a,Timmel1998,Till1998} the mechanism that gives rise to it has yet to be fully explained. Timmel and co-workers have shown that applying a magnetic field lifts degeneracies among the zero field eigenstates of a radical pair with a single proton, and that this changes the time-dependent probability of finding the radical pair in the singlet state (${\rm S}$).\cite{Timmel1998} They also investigated a variety of other radical pairs containing more nuclear spins, with an emphasis on understanding the low field effect on the singlet yield of the recombination reaction.\cite{Timmel1998} However, they did not go on to investigate the more detailed question of how the magnetic field affects the time-dependent probability of finding the radical pair in each of the three triplet states (${\rm T}_+$, ${\rm T}_0$ and ${\rm T}_-$).

Brocklehurst and McLauchlan have attempted to infer the effect of a small magnetic field on these triplet state probabilities from the conservation of angular momentum, again for a radical pair containing just a single proton.\cite{Brocklehurst1996} They argued as follows: (i) The total spin angular momentum quantum number of the radical pair in the singlet electronic state is $J=1/2$, whereas in the triplet state $J$ may be either $1/2$ or $3/2$. (ii) In the absence of a magnetic field, both $J$ and $M_J$ are conserved, but once a field is applied only $M_J$ is conserved. (iii) This implies that the $\ket{J = 1/2, M_J = \pm 1/2} \to \ket{J = 3/2, M_J = \pm 1/2}$ transitions that are forbidden in the absence of the magnetic field become allowed when the field is switched on. (iv) These transitions lead to an increase in ${\rm S} \to {\rm T_\pm}$ interconversion. To quote directly from their paper:\cite{Brocklehurst1996} \lq\lq The field first enables spin evolution of the singlet to at least some of the hyperfine states of the ${\rm T}_{\pm}$ levels to occur".

While much of this argument is correct, we shall show below that the final conclusion in (iv), which Brocklehurst and McLauchlan reached by considering a classical vector model of the electronic and nuclear spin states,\cite{Brocklehurst1996} is not. In fact, the application of a weak magnetic field leads to an increase in ${\rm S}\to {\rm T}_0$ interconversion, and not to an increase in ${\rm S}\to{\rm T}_{\pm}$ interconversion. This distinction is important, because it has implications for the interpretation of the low magnetic field effects in a wide variety of radical pairs. We shall use it here to explain a biphasic-triphasic-biphasic transition with increasing magnetic field strength in the magnetic field effect on the time-dependent survival probability of a photoexcited carotenoid-porphyrin-fullerene (C$^{\cdot +}$PF$^{\cdot -}$) radical pair. Similar behaviour has been observed before for a different radical pair,\cite{Bagryansky2002} but until now remained unexplained.

Sec.~II analyses the low field effect on a radical pair containing a single proton, and on a radical pair in which one of the electrons is hyperfine-coupled to a large number of nuclear spins. The new time-dependent transitions that occur when the field is switched on are shown to be between ${{\rm S}}$ and ${{\rm T}_0}$ in both cases. (The same result is found for a radical pair in which each radical contains a single proton, but since the analysis of this case is much lengthier it is deferred to the supplementary material.) Sec.~III presents experimental measurements and simulations of the magnetic field effect on the transient absorption of the C$^{\cdot +}$PF$^{\cdot -}$ radical pair which support the conclusions drawn in Sec.~II, and Sec.~IV concludes the paper.

\section{Theory}

In the absence of radical pair recombination processes, the time-dependent probability of finding a singlet-born radical pair in the singlet state is given by\cite{Schulten1978,Lewis2014}
\begin{equation}
{\rm P_S}(t)  = {1\over Z}{\rm tr}[\hat{P}_{\rm S} e^{+i\hat{H}t} \hat{P}_{\rm S} e^{-i\hat{H}t}],
\end{equation}
where $Z$ is the total number of nuclear spin states in the radical pair, $\hat{H}$ is its spin Hamiltonian, and
\begin{equation}
\hat{P}_{\rm S} = \ket{{\rm S}}\bra{{\rm S}}
\end{equation}
is the projection operator onto the singlet electronic subspace. (Note that, as is conventional in this field, we are working in a unit system in which $\hbar = 1$.) Rewriting Eq.~(1) in terms of the eigenstates $\ket{n}$ and eigenvalues $\omega_n$ of $\hat{H}$ gives\cite{Timmel1998}
\begin{equation}
{\rm P_S}(t) = {1\over Z}\sum_{m,n} |P^{nm}_{\rm S}|^2 e^{i\omega_{mn}t},
\end{equation}
where $P^{nm}_{\rm S}=\bra{n}\hat{P}_{\rm S}\ket{m}$ and $\omega_{mn} = \omega_m - \omega_n$. Similar expressions are obtained for the probability of finding the radical pair in the ${\rm T}_+$, ${\rm T}_0$, and ${\rm T}_-$ states,
\begin{equation}
\begin{gathered}
{\rm P_{T_+}}(t) = {1\over Z}\sum_{m,n} P^{nm}_{\rm S}P^{mn}_{\rm T_+}  e^{i\omega_{mn}t}, \\
{\rm P_{T_0}}(t) = {1\over Z}\sum_{m,n} P^{nm}_{\rm S}P^{mn}_{\rm T_0}  e^{i\omega_{mn}t}, \\
{\rm P_{T_-}}(t) = {1\over Z}\sum_{m,n} P^{nm}_{\rm S}P^{mn}_{\rm T_-}  e^{i\omega_{mn}t}, 
\end{gathered}
\end{equation}
where $P^{mn}_{\rm T_+}$, $P^{mn}_{\rm T_0}$, and $P^{mn}_{\rm T_-}$ are matrix elements of the projection operators
\begin{equation}
\begin{gathered}
\hat{P}_{\rm T_+} = \ket{{\rm T}_+}\bra{{\rm T}_+}, \\
\hat{P}_{\rm T_0} =  \ket{{\rm T}_0}\bra{{\rm T}_0},\\
\hat{P}_{\rm T_-} = \ket{{\rm T}_-}\bra{{\rm T}_-}.
\end{gathered}
\end{equation}

In order for a term in the double sum over $m$ and $n$ in Eq.~(3) or (4) to depend on time, and hence describe some change in the population of the electronic states of the radical pair, $\omega_{mn}$ must be non-zero. If the application of a weak magnetic field lifts a degeneracy between two eigenstates $m$ and $n$, it could therefore result in an increase in the rate of interconversion between electronic states. However, lifting the degeneracy between states $m$ and $n$ will only affect the time-dependent probability of finding the radical pair in a given electronic state (${\rm S}$, ${\rm T}_+$, ${\rm T}_0$ or ${\rm T}_-$) if the $mn$ matrix element of the operator that projects onto this state is non-zero. 

\subsection{A single nuclear spin in the first radical}

We shall first consider the effect of applying a weak magnetic field to a radical pair with a single $I=1/2$ nuclear spin in one of the radicals and none in the other. This is the problem that was studied by Timmel and co-workers,\cite{Timmel1998} and by Brocklehurst and McLauchlan.\cite{Brocklehurst1996} The spin Hamiltonian is
\begin{equation}
\hat{H} = \omega(\hat{S}_{1z}+\hat{S}_{2z})+a \hat{\bf I} \cdot \hat{\bf S}_1,
\end{equation}
in which $\omega$ is proportional to the applied magnetic field and $a$ is the hyperfine coupling constant between the electron spin and the nuclear spin in the first radical. 

This Hamiltonian has eight eigenstates
\begin{equation}
\begin{gathered}
\ket{1} = \ket{{\rm T}_+\alpha}, \\
\ket{2} = {1\over\sqrt{2}}(\ket{{\rm T}_0\alpha}+\ket{{\rm S}\alpha}), \\
\ket{3} = {c_-\over 2}(\ket{{\rm T}_0\alpha}-\ket{{\rm S}\alpha})+{c_+\over\sqrt{2}}\ket{{\rm T}_+\beta},\\
\ket{4} = {c_+\over 2}(\ket{{\rm T}_0\alpha}-\ket{{\rm S}\alpha})-{c_-\over\sqrt{2}}\ket{{\rm T}_+\beta},\\
\ket{5} = {c_-\over 2}(\ket{{\rm T}_0\beta}+\ket{{\rm S}\beta})-{c_+\over\sqrt{2}}\ket{{\rm T}_-\alpha},\\
\ket{6} = {c_+\over 2}(\ket{{\rm T}_0\beta}+\ket{{\rm S}\beta})+{c_-\over\sqrt{2}}\ket{{\rm T}_-\alpha},\\
\ket{7} = {1\over\sqrt{2}}(\ket{{\rm T}_0\beta}-\ket{{\rm S}\beta}), \\
\ket{8} = \ket{{\rm T}_-\beta}, \\
\end{gathered}
\end{equation}
with eigenvalues
\begin{equation}
\begin{gathered}
\omega_1 = \omega+a/4, \\
\omega_2 = a/4, \\
\omega_3 = (\omega+\mu)/2-a/4, \\
\omega_4 = (\omega-\mu)/2-a/4, \\
\omega_5 = -(\omega+\mu)/2-a/4, \\
\omega_6 = -(\omega-\mu)/2-a/4, \\
\omega_7= a/4, \\
\omega_8 = -\omega+a/4,
\end{gathered}
\end{equation}
where $c_{\pm}=\sqrt{1\pm\omega/\mu}$ and $\mu = \sqrt{\omega^2+a^2}$.

Note that we have sorted these eigenstates in order of decreasing $M_J$: state $\ket{1}$ has $M_J=+3/2$, states $\ket{2}$ to $\ket{4}$ have $M_J=+1/2$, states $\ket{5}$ to $\ket{7}$ have $M_J=-1/2$, and state $\ket{8}$ has $M_J=-3/2$. It follows from points (ii) and (iii) of Brocklehurst and McLauchlan's argument (see Sec.~I) that, as the magnetic field is switched on, the only new time-dependent terms it can introduce into Eqs.~(3) and~(4) are those involving matrix elements between states $\ket{2}$ to $\ket{4}$ and between states $\ket{5}$ to $\ket{7}$. In the first of these groups, the application of the field lifts a zero-field degeneracy between states $\ket{2}$ and $\ket{3}$, and in the second it lifts a degeneracy between states $\ket{6}$ and $\ket{7}$. These are therefore the only pairs of states we need to consider in order to understand the low field effect.

The matrix elements of the operators $\hat{P}_{\rm S}$, $\hat{P}_{{\rm T}_+}$, $\hat{P}_{{\rm T}_0}$ and $\hat{P}_{{\rm T}_-}$ between states $\ket{2}$ and $\ket{3}$ and between states $\ket{6}$ and $\ket{7}$ are as follows:
\begin{equation}
\begin{gathered}
P^{23}_{\rm T_+} = P^{23}_{\rm T_-} = P^{67}_{\rm T_+} = P^{67}_{\rm T_-} = 0, \\
P^{23}_{\rm T_0} = - P^{23}_{\rm S} = {c_-\over 2\sqrt{2}}, \\
P^{67}_{\rm T_0} = - P^{67}_{\rm S} = {c_+\over 2\sqrt{2}}.
\end{gathered}
\end{equation}
The effect of the field in lifting the degeneracies between these eigenstates is therefore to change the time-dependent probabilities of finding the radical pair in the ${\rm S}$ and ${\rm T_0}$ states, by equal amounts and in opposite directions. The symmetry-breaking does not affect the probabilities of finding the radical pair in the ${\rm T_+}$ and ${\rm T_-}$ states. 

\subsection{Many nuclear spins in the first radical}

Applying the same analysis to a radical pair with a very large number of nuclear spins in one of the radicals and none in the other leads to the same conclusion: the new  transitions that occur when the field is switched on are again between ${\rm S}$ and ${\rm T}_0$ and not between ${\rm S}$ and ${\rm T}_{\pm}$. 

In this case, the electron spin dynamics is accurately described by the Schulten-Wolynes approximation,\cite{Schulten1978} in which the nuclear spin operators in the full Hamiltonian
\begin{equation} 
\hat{H} = \omega (\hat{S}_{1z}+\hat{S}_{2z})+\sum_{i=1}^N a_i\hat{\bf I}_i\cdot\hat{\bf S}_1
\end{equation}
are replaced by static hyperfine fields ${\bf h}$ sampled from the Gaussian distribution\cite{Schulten1978,Manolopoulos2013}
\begin{equation}
P({\bf h}) = \left({\tau^2\over 4\pi}\right)^{3/2} e^{-|{\bf h}|^2\tau^2/4},
\end{equation}
where
\begin{equation}
\tau^2 = {6\over \sum_{i=1}^N a_i^2 I_i(I_i+1)}.
\end{equation}

For each ${\bf h}$ sampled from this distribution, the Hamiltonian that governs the electron spin dynamics is simply\cite{Schulten1978}
\begin{equation}
\hat{H} = \omega (\hat{S}_{1z}+\hat{S}_{2z})+{\bf h}\cdot\hat{\bf S}_1,
\end{equation}
in which ${\bf h}$ is a vector rather than a vector operator. This Hamiltonian has four eigenstates
\begin{equation}
\begin{gathered}
\ket{1} = {1\over\sqrt{1+|d_{-}|^2}}\left(\ket{{\rm T}_{-}}-{d_{-}\over\sqrt{2}}\ket{{\rm T}_0}-{d_{-}\over\sqrt{2}}\ket{{\rm S}}\right),\\
\ket{2} = {1\over\sqrt{1+|d_{-}|^2}}\left(d_{-}\ket{{\rm T}_{+}}-{1\over\sqrt{2}}\ket{{\rm T}_0}+{1\over\sqrt{2}}\ket{{\rm S}}\right),\\
\ket{3} = {1\over\sqrt{1+|d_{+}|^2}}\left(\ket{{\rm T}_{-}}+{d_{+}\over\sqrt{2}}\ket{{\rm T}_0}+{d_{+}\over\sqrt{2}}\ket{{\rm S}}\right),\\
\ket{4} = {1\over\sqrt{1+|d_{+}|^2}}\left(d_{+}\ket{{\rm T}_{+}}+{1\over\sqrt{2}}\ket{{\rm T}_0}-{1\over\sqrt{2}}\ket{{\rm S}}\right),\\
\end{gathered}
\end{equation}
with eigenvalues
\begin{equation}
\begin{gathered}
\omega_1 = -{1\over 2}(\omega+\nu),\\
\omega_2 = {1\over 2}(\omega-\nu),\\
\omega_3 = -{1\over 2}(\omega-\nu),\\
\omega_4 = {1\over 2}(\omega+\nu),\\
\end{gathered}
\end{equation}
where $d_{\pm}=[\nu\pm (h_z+\omega)]/[h_x+i h_y]$ and $\nu=\sqrt{h_x^2+h_y^2+(h_z+\omega)^2}$.

In the absence of an external magnetic field, there are two pairs of degenerate eigenstates: states $\ket{1}$ and $\ket{2}$ and states $\ket{3}$ and $\ket{4}$. Both degeneracies are lifted by the application of a field, leading to time-dependent population transfer among the electronic states. The relevant matrix elements of the singlet and triplet projection operators in this case are
\begin{equation}
\begin{gathered}
P^{12}_{\rm T_+} = P^{12}_{\rm T_-} = P^{34}_{\rm T_+} = P^{34}_{\rm T_-} = 0, \\
P^{12}_{\rm T_0} = - P^{12}_{\rm S} = {d_{-}\over 2(1+|d_{-}|^2)} = {{h_x-ih_y}\over 4\nu},\\
P^{34}_{\rm T_0} = - P^{34}_{\rm S} = {d_{+}\over 2(1+|d_{+}|^2)} = {{h_x-ih_y}\over 4\nu}.\\ 
\end{gathered}
\end{equation}
Thus we have the same situation as we had in Sec.~II.A: new time-dependent terms that change the probabilities of finding the radical pair in the ${\rm S}$ and ${\rm T_0}$ states, by equal amounts and in opposite directions, without changing the probabilities of finding the radical pair in the ${\rm T_+}$ and ${\rm T_-}$ states.

\subsection{Singlet and triplet yields}

The effect of radical pair recombination is easiest to analyse in the special case where the singlet and triplet states of the (singlet-born) radical pair have the same recombination rate constant $k_{\rm S}=k_{\rm T}=k$. In this case, the quantum yields of the singlet and triplet products are simply\cite{Timmel1998,Brocklehurst1996,Manolopoulos2013}
\begin{equation}
\Phi_{\rm X} = k\int_0^{\infty} {\rm P}_{\rm X}(t)e^{-kt}\,{\rm d}t,
\end{equation}
for ${\rm X}={\rm S}$, ${\rm T}_+$, ${\rm T}_0$ and ${\rm T}_-$, where ${\rm P}_{\rm X}(t)$ is one of the time-dependent probabilites in Eqs.~(3) and~(4). Substituting these equations into Eq.~(17) and evaluating the time integral gives\cite{Timmel1998}
\begin{equation}
\Phi_{\rm X} = {1\over Z}\sum_{n} P_{\rm S}^{nn}P_{\rm X}^{nn} + {2\over Z}\sum_{m>n} P_{\rm S}^{nm}P_{\rm X}^{mn} {k^2\over k^2+\omega_{mn}^2},
\end{equation}
where we have used the fact that $P_{\rm X}^{nm}=P_{\rm X}^{mn}$ (which holds for all radical pairs with isotropic hyperfine interactions, for which the Hamiltonian matrix is real). 

In the case of a radical pair with a single $I=1/2$ nuclear spin, the matrix elements $P_{\rm X}^{nm}$ are easy to work out from the eigenfunctions in Eq.~(7), allowing all four quantum yields $\Phi_{\rm X}$ to be written in closed form. However, the resulting expressions are rather lengthy so we shall not give them here. Instead, we shall simply illustrate them by plotting the various quantum yields as a function of the applied magnetic field strength $\omega/a$ for a typical problem with $k=a/2$.
 
\begin{figure}[t]
\centering
\resizebox{0.85\columnwidth}{!} {\includegraphics{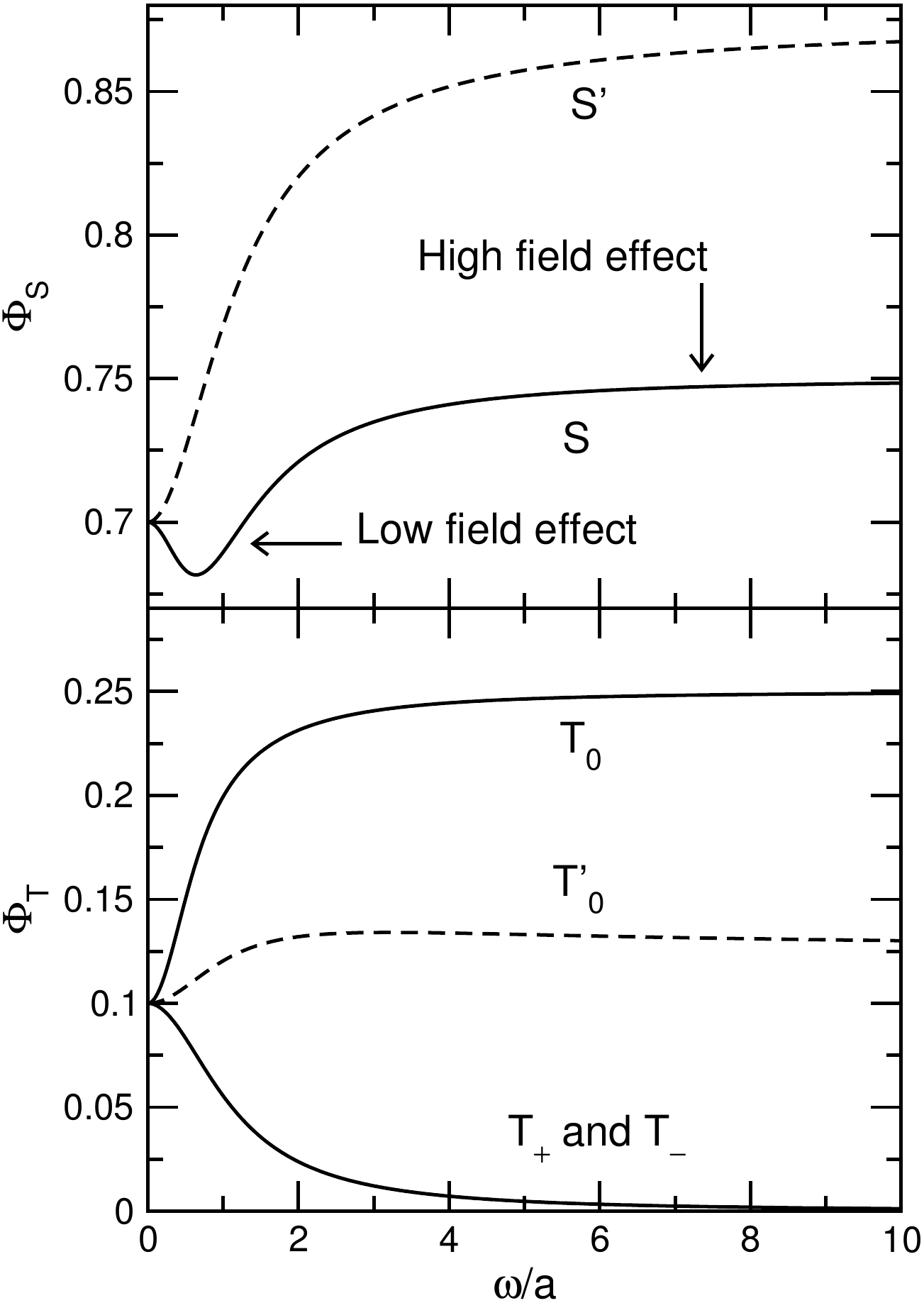}}
\caption{Singlet and triplet yields of a singlet-born radical pair with one proton in one radical and none in the other, as a function of the magnetic field strength $\omega/a$, for $k=a/2$. The dashed curves labelled S$'$ and T$_0'$ are obtained by switching off the magnetic field dependence of the \lq\lq symmetry-breaking" ${\rm S}\to {\rm T}_0$ transitions identified in the text.}
\label{CPFEnergies}
\end{figure}

This plot is shown in Fig.~1, which provides a complete picture of the low and high field effects in this radical pair. The net transfer of population from S to ${\rm T}_0$ increases monotonically with increasing magnetic field strength, and the net transfer of population from S to ${\rm T}_{\pm}$ decreases monotonically. At low field strengths, the increasing ${\rm S}\to{\rm T}_0$ population transfer dominates, leading to a dip in the singlet yield. At high field strengths the decreasing ${\rm S}\to{\rm T}_{\pm}$ population transfer dominates, leading to an increase in the singlet yield. The change in the ${\rm S}\to{\rm T}_0$ population transfer is due both to the symmetry-breaking effect identified above and to the dependence of the eigenstates and eigenvalues of $\hat{H}$ on $\omega$, whereas the change in the ${\rm S}\to{\rm T}_{\pm}$ population transfer is solely due to the $\omega$-dependence of the eigenstates and eigenvalues. Note that the ${\rm T}_{\pm}$ yields go to zero as the Zeeman splitting of these states increases and they become energetically inaccessible from the singlet state: this is the well known (and well understood) high field effect.\cite{Brocklehurst1976,Werner1978,Weller1983}

The low field effect in Fig.~1 -- the dip in the singlet yield at low magnetic field strengths -- is more difficult to understand because it arises from a competition between increasing ${\rm S}\to {\rm T}_0$ population transfer and decreasing ${\rm S}\to {\rm T}_{\pm}$ population transfer with increasing magnetic field strength. However, the dashed curves labelled ${\rm S}'$ and ${\rm T}_0'$ in the figure make it clear that the low field effect would not be observed without the symmetry-breaking ${\rm S}\to {\rm T}_0$ transitions identified above. These curves are obtained when the magnetic field dependence of these transitions is artificially switched off by replacing the corresponding terms in Eq.~(18) with their zero-field values. This reduces the increase in ${\rm S}\to {\rm T}_0$ population transfer with increasing magnetic field strength and eliminates the low field effect entirely. Thus the low field effect can unambiguously be attributed to the symmetry-breaking ${\rm S}\to {\rm T}_0$ transitions.

\begin{figure*}[t]
\centering
\resizebox{1.6\columnwidth}{!} {\includegraphics{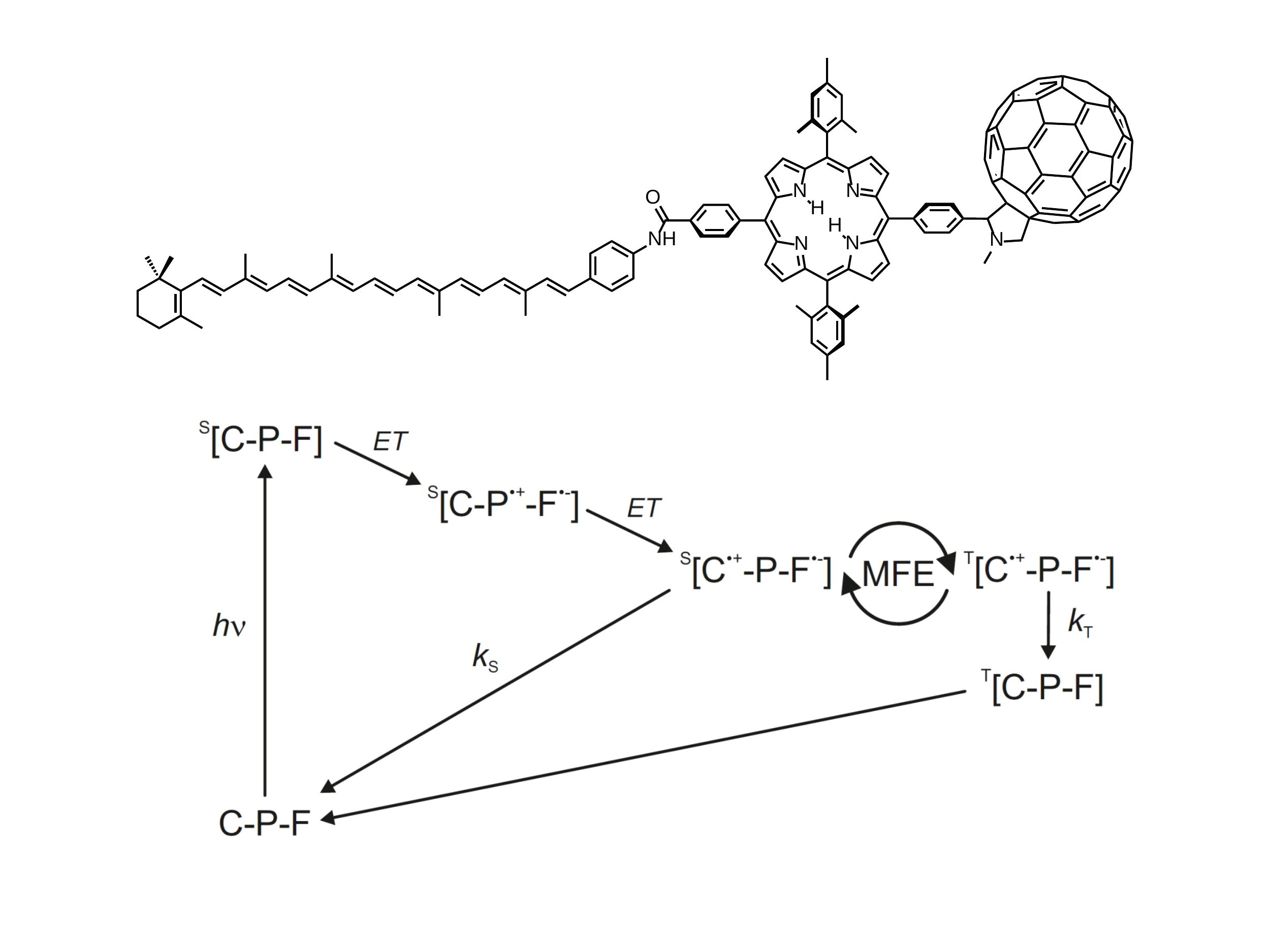}}
\caption{The carotenoid-porphyrin-fullerene triad, along with a simplified diagram showing the photochemistry \cite{Kodis04} that precedes the coherent electron spin evolution and asymmetric ($k_{\rm S}\gg k_{\rm T}$) recombination of the C$^{\cdot +}$PF$^{\cdot -}$ radical pair. For more detail on the photochemistry and recombination characteristics, see Maeda {\em et. al.}\cite{Maeda2008}}
\label{triad}
\end{figure*}

\section{Example Application}

Experimentally, it is rarely possible to probe the individual triplet state populations ${\rm P_{T_+}}(t)$, ${\rm P_{T_0}}(t)$ and ${\rm P_{T_-}}(t)$. However, the effect of an applied magnetic field on these populations does have implications for the interpretation of various other experimental observables, as we shall now illustrate with a combined experimental and theoretical study of a C$^{\cdot +}$PF$^{\cdot -}$ radical pair.

The CPF triad is shown in Fig.~2, which also contains a summary of its photochemistry.\cite{Kodis04,Maeda2008} After photoexcitation of the porphyrin chromophore two rapid consecutive electron transfer steps produce the C$^{\cdot +}$PF$^{\cdot -}$ radical pair, predominantly in its singlet state.\cite{Maeda2011, Kodis04} The radical pair then undergoes coherent spin evolution between its singlet and triplet states, which decay with different rate constants $k_{\rm S}\not= k_{\rm T}$ to different products.\cite{Maeda2011} Because these rate constants are different, the lifetime of the radical pair depends on the extent of the singlet to triplet intersystem crossing, which depends in turn on the strength of the applied magnetic field.\cite{Maeda2008,Maeda2011} The radical pair survival probability can be determined in the presence and the absence of a magnetic field by measuring the transient absorption of the carotenoid radial C$^{\cdot +}$ at different times after the initial photoexcitation laser pulse. In what follows, we shall provide a detailed explanation for the magnetic field effects (MFEs) that emerge from these measurements in terms of the competition between ${\rm S}\to {\rm T_0}$ and ${\rm S}\to {\rm T_{\pm}}$ transitions, and show that this explanation is entirely consistent with the conclusions we have drawn above.

\subsection{Experimental details}

The experimental setup and procedures used to measure the transient absorption of the C$^{\cdot +}$PF$^{\cdot -}$ radical pair are broadly similar to those described in Ref.~\onlinecite{Maeda2008}. A full technical account of the experimental details is beyond the scope of this paper  and will be published elsewhere. Briefly, the CPF triad molecules are excited using a 532\,nm Nd:YAG laser with 10\,Hz repetition rate and 7\,ns pulse duration, at pump energies of $<$ 1\,mJ. The total transient radical pair population at time $t$ after the initial photoexcitation laser pulse is probed via the absorption of the carotenoid radical cation at 980\,nm. The difference in the sample absorption in the presence and absence of the pump pulse is labelled $\Delta$A$(t)$. 

\begin{figure*}[t]
\centering
\resizebox{1.4\columnwidth}{!} {\includegraphics{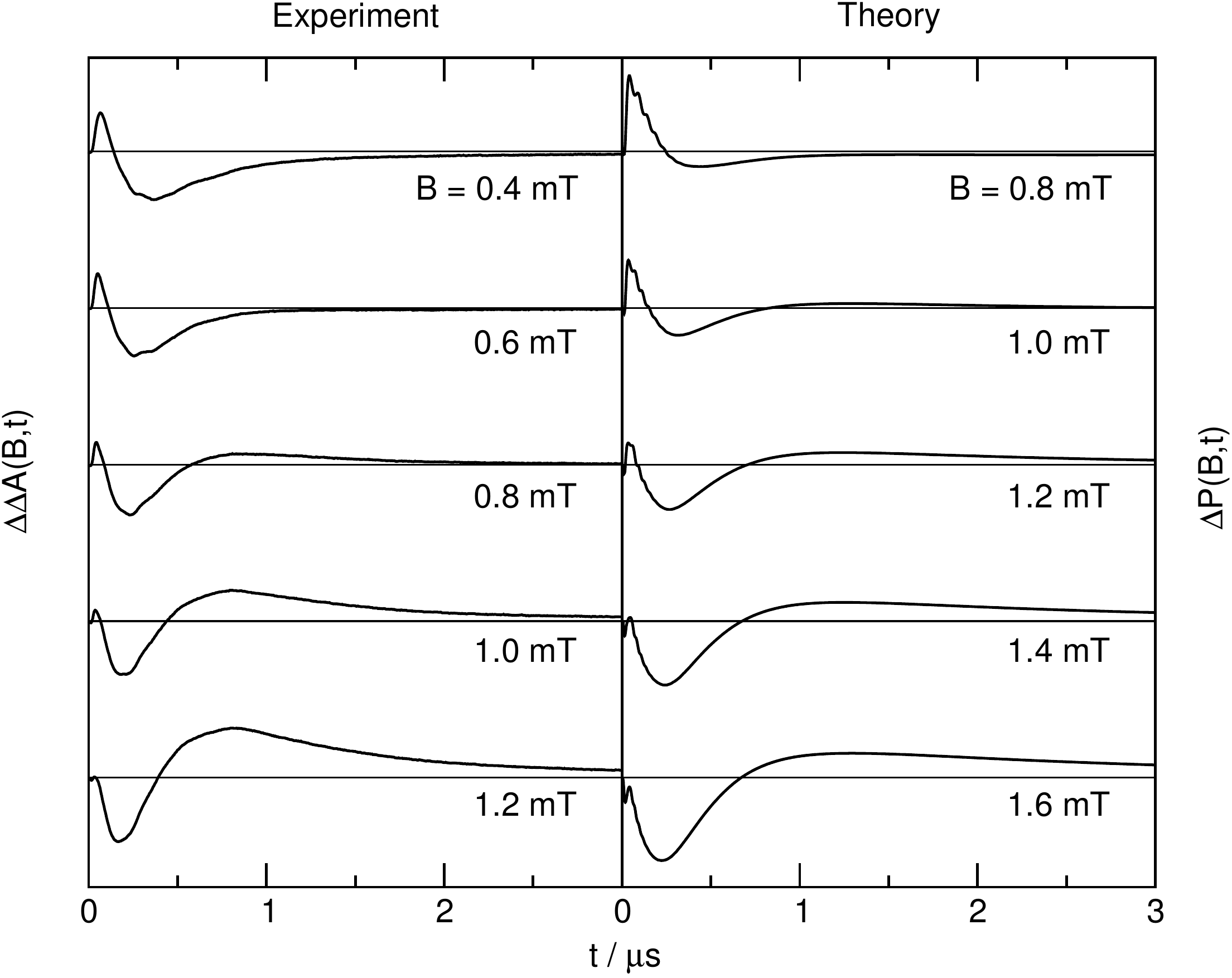}}
\caption{Left: the magnetic field effect $\Delta \Delta$A$(B,t)$ in the experimentally measured absorption of the C$^{\cdot +}$PF$^{\cdot -}$ radical pair over a range of applied magnetic field strengths. Right: the magnetic field effect $\Delta {\rm P}(B,t)$ in the semiclassically simulated survival probability of the radical pair over a slightly higher range of field strengths.}
\end{figure*}

Magnetic field effects are measured by comparing $\Delta$A$(t)$ recorded in the presence and absence of an external magnetic field; the experimental signals are thus of the type $\Delta \Delta$A$(B,t)$ = $\Delta$A$(B,t)-\Delta$A$(0,t)$. The magnetic fields are created by three sets of orthogonal Helmholtz coils, positioned around the sample cell and the cryostat. Careful calibration of all coils allows cancellation of the Earth's magnetic field. The experiments are performed in 2-methyltetrahydrofuran at 120 K.

The results of these transient absorption experiments are shown in the left-hand panels of Fig.~3. As in previous studies of the CPF triad, biphasic behaviour is observed in both the low field region (below 0.6\,mT) and the high field region (above 1.2\,mT)\cite{Maeda2008}. However, at intermediate field strengths (between 0.8 and 1.0\,mT), triphasic behaviour is observed, with a positive MFE at both very short times and long times, and a negative MFE at intermediate times. This behaviour has been observed before for a different radical pair,\cite{Bagryansky2002} but not explained in terms of the magnetic field and time dependence of ${\rm S}\to{\rm T}_0$ and ${\rm S}\to{\rm T}_{\pm}$ population transfer as we shall do here.

\subsection{Computational details}

In order to understand the origin of the triphasic behaviour, we have used the semiclassical method described in Ref.~\onlinecite{Lewis2014} to calculate the magnetic field effect on the survival probability of the C$^{\cdot +}$PF$^{\cdot -}$ radical pair. This is defined as $\Delta {\rm P}(B,t) = {\rm P}(B,t)-{\rm P}(0,t)$, where ${\rm P}(B,t)$ is the sum of singlet and triplet populations of the radical pair at time $t$ after the initial photoexcitation laser pulse, in a magnetic field of strength $B$. The radical pair recombination processes that deplete the singlet and triplet populations are included in the calculation of ${\rm P}(B,t)$ with a Haberkorn recombination operator\cite{Haberkorn1976} so as to give a $\Delta {\rm P}(B,t)$ proportional to the experimental signal $\Delta\Delta$A$(B,t)$. The semiclassical calculation of ${\rm P}(B,t)$ is described in detail in Ref.~\onlinecite{Lewis2014}, where it is shown to reproduce the experimentally measured effect of an Earth-strength ($\sim$\,50\,$\mu$T) magnetic field on the transient absorption of the radical pair.\cite{Maeda2008}

The input to the semiclassical calculation consists of the hyperfine coupling constants of the C$^{\cdot +}$ and F$^{\cdot -}$ radicals and the recombination rate constants for the singlet and triplet states of the radical pair. The latter have been inferred from EPR experiments\cite{Maeda2011} to be $k_{\rm S} = 1.8\times 10^7$ s$^{-1}$ and $k_{\rm T} = 7.1\times 10^{4}$ s$^{-1}$ at 110\,K. We have used these values in our calculations even though the present experiments were performed at 120\,K. The isotropic hyperfine coupling constants of the 45 hydrogen nuclei on the carotenoid radical have been calculated using B3LYP density functional theory\cite{Lee1988,Becke1993} with the EPR-II basis set,\cite{Barone1995} and are given in the Appendix of Ref.~\onlinecite{Lewis2014}. $^{13}$C nuclei will be present with $\sim 1$\% natural abundance in both radicals, but we have neglected the hyperfine coupling to these nuclei in our calculations. 

The semiclassically computed $\Delta {\rm P}(B,t)$ with these parameters is shown on the right-hand side of Fig.~3. Our simulations agree qualitatively with the experimental data, with the MFE exhibiting biphasic behaviour with reversed phases in the low and high field regions, and triphasic behaviour at intermediate field strengths. The agreement is not quantitative: the biphasic--triphasic and triphasic--biphasic transitions occur at higher field strengths in the simulations than in the experiment. There could be several reasons for this, including inaccuracies in our values for the recombination rate constants and hyperfine coupling constants, the presence of a small ($\sim 7\%$) fraction of triplet-born radical pairs in the experiment,\cite{Lewis2014,Maeda2011} the neglect of any electron spin coupling or relaxation in the calculations,\cite{Schulten1978,Steiner1989} and errors in the semiclassical approximation itself.\cite{Manolopoulos2013,Lewis2014,Lewis2016,Lindoy2018} However, the agreement is certainly good enough for us to use  our simulations to shed light on the origin of the biphasic--triphasic-biphasic transition in the experimental signal, which we shall do next.

\subsection{Discussion of the results}

\begin{figure}[t]
\centering
\resizebox{0.9\columnwidth}{!} {\includegraphics{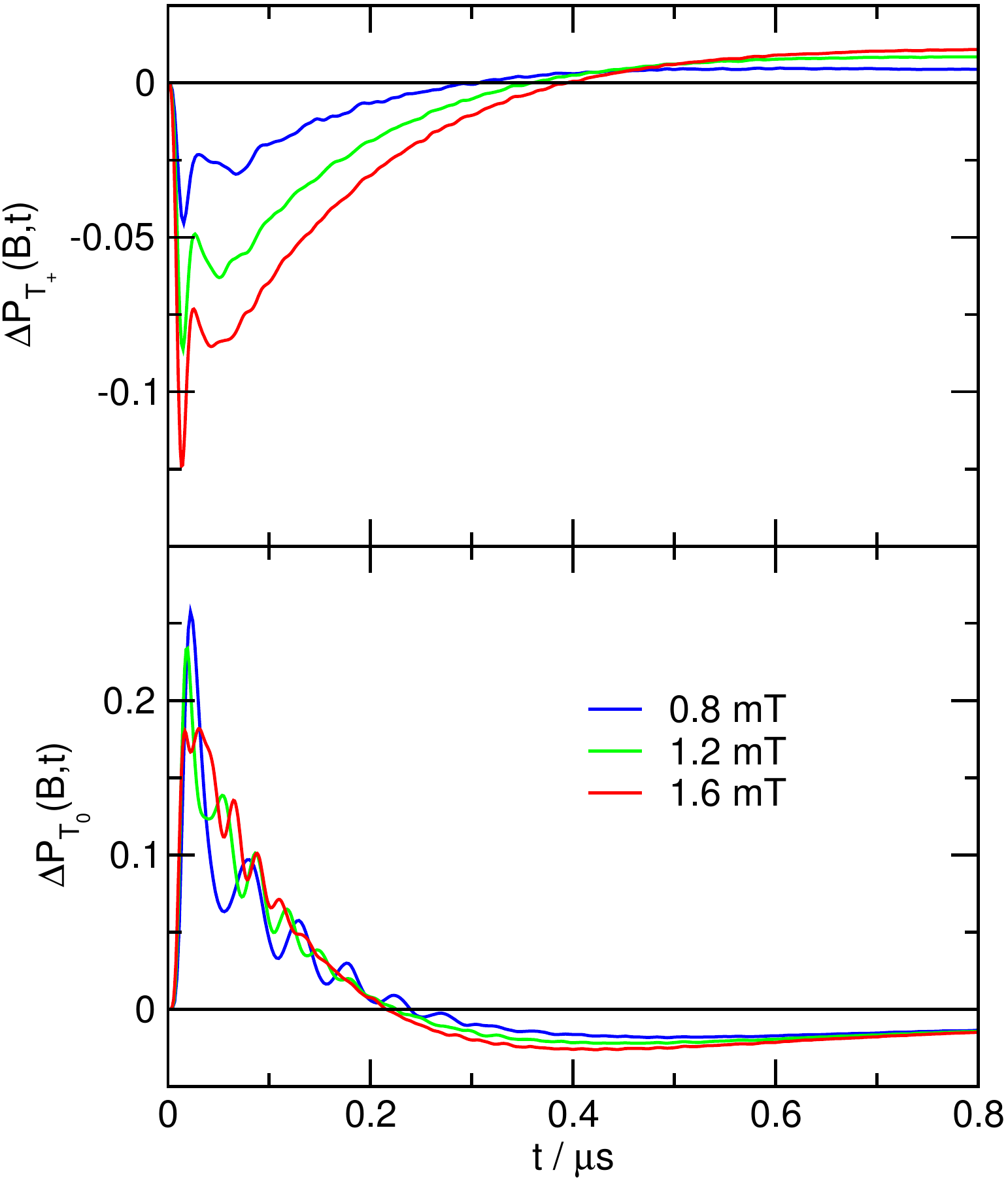}}
\caption{Magnetic field effects $\Delta {\rm P_{X}}(B,t)$ in the time-dependent populations of the ${\rm X}={\rm T_+}$ (top) and ${\rm X}={\rm T_0}$ (bottom) states of the C$^{\cdot +}$PF$^{\cdot -}$ radical pair at three different magnetic field strengths, from the present semiclassical calculations.}
\end{figure}

The triphasic behaviour of the magnetic field effect can be explained by examining how the probability of finding the radical pair in the ${{\rm T}_+}$, ${{\rm T}_-}$, and ${{\rm T}_0}$ states changes with the magnetic field strength. The magnetic field dependence of the probability of being in the ${{\rm T}_+}$ state, $\Delta {\rm P_{T_+}}(B,t)={\rm P_{T_+}}(B,t)-{\rm P_{T_+}}(0,t)$, is shown in Fig.~4 (top panel). This is nearly identical to $\Delta {\rm P_{T_-}}(B,t)$, which is not shown. At short times, $\Delta {\rm P_{T_+}}(B,t)$ decreases with increasing $B$, and at long times it increases with increasing $B$. This is to be expected from the high field effect.\cite{Brocklehurst1976} As $B$ increases, the ${\rm T_{\pm}}$ states become energetically separated from the S state, which reduces the extent of ${\rm S}\to {\rm T}_{\pm}$ intersystem crossing.  At early times, less of the singlet-born radical pair is transferred to the ${\rm T_+}$ state as $B$ increases, resulting in a dip in $\Delta {\rm P_{T_+}}(B,t)$ that becomes more pronounced with increasing $B$. And at late times, less of the ${\rm T}_+$ population is transferred back to S as $B$ increases, resulting in a peak in $\Delta {\rm P_{T_+}}(B,t)$ that increases with increasing $B$. 

The behaviour of $\Delta {\rm P_{T_0}}(B,t) = {\rm P_{T_0}}(B,t)-{\rm P_{T_0}}(0,t)$ shown in Fig.~4 (bottom panel) is quite the opposite, as would be expected from the analysis in Sec.~II. This exhibits a peak at short times and a dip at long times, consistent with {\em more} ${\rm S}\to {\rm T}_0$ population transfer in the presence than the absence of the field. The short-time $\Delta {\rm P_{T_0}}(B,t)$ peak exhibits a clear Rabi oscillation, the frequency of which is consistent with hyperfine-mediated ${\rm S}\leftrightarrow {\rm T}_0$ interconversion in the Zeeman field, and the long-time dip becomes more pronounced with increasing $B$ as a result of more efficient long-time ${\rm T}_0\to {\rm S}$ back transfer at higher field strengths. All of this is clearly consistent with the analysis in Sec.~II and with what one would expect from the ${\rm T}_0$ and ${\rm T}_{\pm}$ quantum yields in Fig.~1.

Taken together, these observations explain the biphasic--triphasic--biphasic behaviour observed in Fig.~3. At low fields, where the presence of the magnetic field only makes a small difference to ${\rm P_{T_+}}(t)$ and ${\rm P_{T_-}}(t)$, the low field effect in ${\rm P_{T_0}}(t)$ dominates the MFE in the survival probability, which is positive at short times and negative at longer times. The reverse is true at high fields, where the high field effect in ${\rm P_{T_+}}(t)$ and ${\rm P_{T_-}}(t)$ outweighs the low field effect in ${\rm P_{T_0}}(t)$. At intermediate field strengths the competition between the two effects results in the observed triphasic behaviour, because the MFEs on ${\rm P_{T_0}}(t)$ and ${\rm P_{T_\pm}}(t)$ are in opposite directions and have different timescales. 

\section{Concluding Remarks}

In this paper, we have shown that the application of a weak magnetic field leads to new pathways for time-dependent population transfer between the S and ${\rm T}_0$ states of a radical pair, and not between the S and ${\rm T}_{\pm}$ states as has previously been claimed.\cite{Brocklehurst1996} Since this result holds for a radical pair with a large number of hyperfine-coupled nuclear spins in one of the two radicals, in the opposite limit of a radical pair with just a single nuclear spin, and in the case of a radical pair in which each radical contains a single nuclear spin (see the supplementary material), we believe it to be a completely general explanation for what has come to be known as the \lq\lq low field effect" in radical pair reactions. We have also illustrated the result by showing how it accounts, when combined with the well known and less controversial \lq\lq high field effect", for the unusual triphasic behaviour observed at certain magnetic field strengths in experimental measurements of the transient absorption of a photoexcited carotenoid-porphyrin-fullerene radical pair. 

\section{Supplementary Material}

The supplementary material presents an analysis of a radical pair in which each radical contains a single proton, and shows that the new transitions that occur in the presence of a magnetic field are again between S and T$_0$ and not between S and T$_{\pm}$. It also presents some additional numerical calculations which demonstrate the robustness of our conclusions about the low field effect to the presence of an exchange coupling between the electron spins in the radical pair.

\begin{acknowledgements}
We would like to thank Devens Gust for the synthesis of the experimental model system and many useful discussions on the triad's photochemistry, and Peter Hore for helpful comments on the first draft of this manuscript. Thomas Fay is supported by Clarendon and E. A. Haigh Scholarships and by the EPSRC Centre for Doctoral Training in Theory and Modelling in the Chemical Sciences, EPSRC grant no. EP/L015722/1. Christian Kerpal and Christiane Timmel are grateful for funding from the US Air Force (USAF) Office of Scientific Research (Air Force Materiel Command, USAF Award FA9550-14-1-0095). Christian Kerpal thanks the DFG for a Research Fellowship (Projekt number 256837888).  Financial support from the EPSRC (EPL011972/1, EP/M016110/1 and EP/J015067/1) and the ERC (grant 320969) is also gratefully acknowledged. Alan Lewis is the leading theoretical author, and Christian Kerpal the leading experimental author, of this paper.

\end{acknowledgements}

\end{document}


\def\bra#1{\left<{#1}\right|}
\def\ket#1{\left|{#1}\right>}
\def\expval#1#2{\bra{#2} {#1} \ket{#2}}
\def\mapright#1{\smash{\mathop{\longrightarrow}\limits^{_{_{\phantom{X}}}{#1}_{_{\phantom{X}}}}}}

\title{On the low magnetic field effect in radical pair reactions:\\ Supplementary material}

\author{Alan M.~Lewis}
\affiliation{Department of Chemistry, University of Oxford, Physical and Theoretical Chemistry Laboratory, South Parks Road, Oxford, OX1 3QZ, UK}

\author{Thomas P. Fay}
\affiliation{Department of Chemistry, University of Oxford, Physical and Theoretical Chemistry Laboratory, South Parks Road, Oxford, OX1 3QZ, UK}

\author{David E.~Manolopoulos}
\affiliation{Department of Chemistry, University of Oxford, Physical and Theoretical Chemistry Laboratory, South Parks Road, Oxford, OX1 3QZ, UK}

\author{Christian Kerpal}
\affiliation{Department of Chemistry, University of Oxford, Inorganic\\ Chemistry Laboratory, South Parks Road, Oxford, OX1 3QR, UK}

\author{Sabine Richert}
\affiliation{Department of Chemistry, University of Oxford, Inorganic\\ Chemistry Laboratory, South Parks Road, Oxford, OX1 3QR, UK}

\author{Christiane R.~Timmel}
\affiliation{Department of Chemistry, University of Oxford, Inorganic\\ Chemistry Laboratory, South Parks Road, Oxford, OX1 3QR, UK}

\begin{abstract}
In the main text of our paper, we have shown that the application of an external magnetic field induces new pathways for ${\rm S}\to{\rm T}_0$ population transfer (and no new pathways for ${\rm S}\to{\rm T}_{\pm}$ population transfer) for two particular radical pairs: one with a single proton in the first radical and no magnetic nuclei in the other (Sec.~II.A), and another with many magnetic nuclei in the first radical and none in the other (Sec.~II.B). 

The main question that this leaves unanswered is whether or not the ${\rm S}\to{\rm T}_0$ result only holds when one of the two radicals does not contain any hyperfine-coupled nuclear spins. Here we investigate this question by considering the case in which each radical is hyperfine-coupled to a single $I=1/2$ nuclear spin. The algebra is lengthier in this case than in either of the cases we have presented in the text, which is why we have deferred it to supplementary material. 

The upshot is that the result still holds: when each radical contains a single hyperfine-coupled nuclear spin, an applied magnetic field leads to new ${\rm S}\to{\rm T}_0$  transitions and to no new ${\rm S}\to{\rm T}_{\pm}$ transitions.

Also, in the main text of our paper we assume that there is no exchange coupling between the electron spins. Exchange coupling lifts the zero-field degeneracy of the singlet and triplet states, however the application of a weak magnetic field still leads to enhancement of ${\rm S}\to {\rm T}_0$ transitions, as well as an enhancement in ${\rm S} \to {\rm T}_+$ transitions due decreasing the energy gap between these two states.

\end{abstract}

\maketitle

\newcommand{\pP}{\ensuremath{\mathcal{P}}}
\newcommand{\pQ}{\ensuremath{\mathcal{Q}}}
\newcommand{\pL}{\ensuremath{\mathcal{L}}}
\newcommand{\pK}{\ensuremath{\mathcal{K}}}
\newcommand{\pG}{\ensuremath{\mathcal{G}}}
\newcommand{\sys}{\ensuremath{\mathrm{s}}}
\newcommand{\nuc}{\ensuremath{\mathrm{n}}}
\newcommand{\el}{\ensuremath{\mathrm{e}}}
\newcommand{\eq}{\ensuremath{\mathrm{eq}}}
\newcommand{\sing}{\ensuremath{\mathrm{S}}}
\newcommand{\trip}{\ensuremath{\mathrm{T}}}
\newcommand{\ipi}{\ensuremath{\mathrm{I}}}
\renewcommand{\op}[1]{\ensuremath{{#1}}}
\newcommand{\nint}[1]{\ensuremath{{#1}^\nuc}}
\newcommand{\ham}{\hat{H}}
\newcommand{\spin}{\hat{\vb{S}}}
\newcommand{\spn}{\hat{S}}
\newcommand{\ispin}{\hat{\vb{I}}}
\newcommand{\ispn}{\hat{I}}

\section{One nuclear spin in each radical}

Consider a radical pair with a single $I=1/2$ nuclear spin in each radical. The Hamiltonian is
\begin{align}
	\ham = \ham_1 + \ham_2, \nonumber
\end{align}
where
\begin{align}
\ham_i = \omega \spn_{iz} + a_i \spin_i\cdot \ispin_i. \nonumber
\end{align}
First diagonalising $\ham_i$ we obtain the following four eigenstates and eigenvalues,
\begin{align}
\begin{array}{ll}
	\ket{1_i} = \ket{\alpha_{S_i}\alpha_{I_i}}, &\quad \omega_{1,i}=a_i/4+\omega/2, \\
	\ket{2_i} = c_{i+}\ket{\alpha_{S_i}\beta_{I_i}} + c_{i-}\ket{\beta_{S_i}\alpha_{I_i}}, &\quad \omega_{2,i}=-a_i/4+\mu_i/2, \\
	\ket{3_i} = c_{i-}\ket{\alpha_{S_i}\beta_{I_i}} - c_{i+}\ket{\beta_{S_i}\alpha_{I_i}},&\quad \omega_{3,i}=-a_i/4-\mu_i/2, \\
	\ket{4_i} = \ket{\beta_{S_i}\beta_{I_i}},&\quad \omega_{4,i}=a_i/4-\omega/2, \nonumber
\end{array}
\end{align}
where
\begin{align}
\mu_i = \sqrt{\omega^2+a_i^2}, \nonumber
\end{align}
and 
\begin{align}
c_{i\pm} &= \sqrt{\frac{\mu_i\pm\omega}{2\mu_i}}. \nonumber
\end{align}

Overall there are 16 eigenstates of the full Hamiltonian. These states $\ket{k}$ are direct products of the $\ket{n_1}$ and $\ket{m_2}$ eigenstates of $\hat{H}_1$ and $\hat{H}_2$, $\ket{n_1,m_2}$. The eigenstates of $\hat{H}$ are also eigenstates of $\hat{J}_z=\hat{S}_{1z}+\hat{I}_{1z}+\hat{S}_{2z}+\hat{I}_{2z}$ and they can be divided into sets with the same eigenvalue $M_J$.  The eigenstate with $M_J=2$ is
\begin{subequations}
\begin{align*}
\ket{1} &= \ket{1_1, 1_2} = \ket{\trip_+\alpha\alpha}.
\end{align*}
i.e., the direct product of the coupled electron spin state $\ket{{\rm T}_+}=\ket{\alpha_{S_1}\alpha_{S_2}}$ with the nuclear spin state
$\ket{\alpha_{I_1}\alpha_{I_2}}$. In the same notation, the eigenstates with $M_J=1$ are
\begin{align*}
\ket{2} &= \ket{1_1,2_2} = c_{2+}\ket{\trip_+\alpha\beta} + \frac{c_{2-}}{\sqrt{2}}\left(\ket{\trip_0\alpha\alpha}+\ket{\sing\alpha\alpha}\right),\\
\ket{3} &= \ket{1_1,3_2}=c_{2-}\ket{\trip_+\alpha\beta} - \frac{c_{2+}}{\sqrt{2}}\left(\ket{\trip_0\alpha\alpha}+\ket{\sing\alpha\alpha}\right),\\
\ket{4} &= \ket{2_1,1_2}= c_{1+}\ket{\trip_+\beta\alpha} + \frac{c_{1-}}{\sqrt{2}}\left(\ket{\trip_0\alpha\alpha}-\ket{\sing\alpha\alpha}\right),\\
\ket{5} &= \ket{3_1,1_2} = c_{1-}\ket{\trip_+\beta\alpha} - \frac{c_{1+}}{\sqrt{2}}\left(\ket{\trip_0\alpha\alpha}-\ket{\sing\alpha\alpha}\right),
\end{align*}
the eigenstates with $M_J = 0$ are
\begin{align*}
\ket{6} &= \ket{1_1,4_2}=\frac{1}{\sqrt{2}}\left(\ket{\trip_0\alpha\beta}+\ket{\sing\alpha\beta}\right),\\
\ket{7} &= \ket{2_1,2_2}=c_{1+}c_{2+}\ket{\trip_+\beta\beta} + \frac{c_{1+}c_{2-}}{\sqrt{2}}\left(\ket{\trip_0\beta\alpha}+\ket{\sing\beta\alpha}\right)\\
&+\frac{c_{1-}c_{2+}}{\sqrt{2}}\left(\ket{\trip_0\alpha\beta}-\ket{\sing\alpha\beta}\right)+c_{1-}c_{2-}\ket{\trip_-\alpha\alpha},  \\
\ket{8} &=\ket{2_1,3_2}=c_{1+}c_{2-}\ket{\trip_+\beta\beta} - \frac{c_{1+}c_{2+}}{\sqrt{2}}\left(\ket{\trip_0\beta\alpha}+\ket{\sing\beta\alpha}\right)\\
&+\frac{c_{1-}c_{2-}}{\sqrt{2}}\left(\ket{\trip_0\alpha\beta}-\ket{\sing\alpha\beta}\right)-c_{1-}c_{2+}\ket{\trip_-\alpha\alpha} , \\ 
\ket{9} &= \ket{3_1,2_2}=c_{1-}c_{2+}\ket{\trip_+\beta\beta} +\frac{c_{1-}c_{2-}}{\sqrt{2}}\left(\ket{\trip_0\beta\alpha}+\ket{\sing\beta\alpha}\right)\\
&-\frac{c_{1+}c_{2+}}{\sqrt{2}}\left(\ket{\trip_0\alpha\beta}-\ket{\sing\alpha\beta}\right)-c_{1+}c_{2-}\ket{\trip_-\alpha\alpha},  \\
\ket{10} &=\ket{3_1,3_2}=c_{1-}c_{2-}\ket{\trip_+\beta\beta} -\frac{c_{1-}c_{2+}}{\sqrt{2}}\left(\ket{\trip_0\beta\alpha}+\ket{\sing\beta\alpha}\right)\\
&-\frac{c_{1+}c_{2-}}{\sqrt{2}}\left(\ket{\trip_0\alpha\beta}-\ket{\sing\alpha\beta}\right)+c_{1+}c_{2+}\ket{\trip_-\alpha\alpha}, \\
\ket{11} &= \ket{4_1,1_2}=\frac{1}{\sqrt{2}}\left(\ket{\trip_0\beta\alpha}-\ket{\sing\beta\alpha}\right),
\end{align*}
the eigenstates with $M_J=-1$ are
\begin{align*}
\ket{12} &= \ket{4_1,2_2} = \frac{c_{2+}}{\sqrt{2}}\left(\ket{\trip_0\beta\beta}-\ket{\sing\beta\beta}\right) + c_{2-}\ket{\trip_-\beta\alpha},\\
\ket{13} &= \ket{4_1,3_2} = \frac{c_{2-}}{\sqrt{2}}\left(\ket{\trip_0\beta\beta}-\ket{\sing\beta\beta}\right) - c_{2+}\ket{\trip_-\beta\alpha},\\
\ket{14} &= \ket{2_1,4_2} = \frac{c_{1+}}{\sqrt{2}}\left(\ket{\trip_0\beta\beta}+\ket{\sing\beta\beta}\right) + c_{1-}\ket{\trip_-\alpha\beta},\\
\ket{15} &= \ket{3_1,4_2} = \frac{c_{1-}}{\sqrt{2}}\left(\ket{\trip_0\beta\beta}+\ket{\sing\beta\beta}\right) - c_{1+}\ket{\trip_-\alpha\beta},
\end{align*}
and the eigenstate with $M_J=-2$ is
\begin{align}
\ket{16} &= \ket{4_1,4_2} = \ket{\trip_- \beta \beta}. \nonumber 
\end{align}
\end{subequations}
The eigenvalues of these states are
\begin{subequations}
\begin{align*}
\omega_1 &= \omega + \frac{a_1+a_2}{4},\\
\omega_2 &= \frac{\omega+\mu_2}{2} + \frac{a_1-a_2}{4},\\
\omega_3 &= \frac{\omega-\mu_2}{2} + \frac{a_1-a_2}{4}, \\
\omega_4 &= \frac{\omega+\mu_1}{2} + \frac{-a_1+a_2}{4},\\
\omega_5 &= \frac{\omega-\mu_1}{2} + \frac{-a_1+a_2}{4}, \\
\omega_6 &= \frac{a_1+a_2}{4}, \\
\omega_7 &= -\frac{a_1+a_2}{4}+\frac{\mu_1+\mu_2}{2}, \\
\omega_8 &= -\frac{a_1+a_2}{4}+\frac{\mu_1-\mu_2}{2}, \\
\omega_9 &= -\frac{a_1+a_2}{4}-\frac{\mu_1-\mu_2}{2}, \\
\omega_{10} &= -\frac{a_1+a_2}{4}-\frac{\mu_1+\mu_2}{2}, \\
\omega_{11} &= \frac{a_1+a_2}{4}, \\
\omega_{12} &= -\frac{\omega-\mu_2}{2} + \frac{a_1 - a_2}{4}, \\
\omega_{13} &= -\frac{\omega+\mu_2}{2} + \frac{a_1 - a_2}{4}, \\
\omega_{14} &= -\frac{\omega-\mu_1}{2} - \frac{a_1 - a_2}{4}, \\
\omega_{15} &= -\frac{\omega+\mu_1}{2} - \frac{a_1 - a_2}{4}, \\
\omega_{16} &= -\omega + \frac{a_1+a_2}{4}. \\
\end{align*}
\end{subequations}
Transitions can only occur between states with the same $M_J$ eigenvalue. In the $M_J=1$ block states $\ket{2}$ and $\ket{4}$ are degenerate when $\omega=0$. This degeneracy is lifted when $\omega\neq0$. In the $M_J=0$ set, states $\ket{6}, \ket{7}$ and $\ket{11}$ are degenerate at zero field. The degeneracies between $\ket{6}$ and $\ket{7}$, and between states $\ket{7}$ and $\ket{11}$, are lifted when $\omega\neq 0$. In the $M_J=-1$ block degeneracies between states $\ket{12}$ and $\ket{14}$ are lifted when $\omega\neq 0$. The matrix elements of the $\hat{P}_{\trip_\pm}$, $\hat{P}_{\trip_0}$ and $\hat{P}_\sing$ operators between each pair of states for which the degeneracy is lifted by the application of a field  are as follows:
\begin{align}
P_{\trip_\pm}^{2,4} = &\ P_{\trip_\pm}^{6,7} = P_{\trip_\pm}^{7,11} = P_{\trip_\pm}^{12,14} = 0 \nonumber \\
P_{\trip_0}^{2,4} &= -P_\sing^{2,4} =  \frac{1}{2}c_{1-}c_{2-} \nonumber\\
P_{\trip_0}^{6,7} &= -P_\sing^{6,7} =  \frac{1}{2}c_{1-}c_{2+} \nonumber \\
P_{\trip_0}^{7,11} &= -P_\sing^{7,11} = \frac{1}{2}c_{1+}c_{2-} \nonumber\\
P_{\trip_0}^{12,14} &= -P_\sing^{12,14} = \frac{1}{2}c_{1+}c_{2+}. \nonumber
\end{align}
As for the one proton radical pair considered in the main text, we see that when degeneracies of radical pair eigenstates are lifted by the application of a magnetic field, there are no new transitions between $\sing$ and $\trip_\pm$ states, but new transitions between $\sing$ and $\trip_0$ states are introduced. 

\section{The Effect of Exchange Coupling}

Exchange coupling lifts the zero-field degeneracy of the singlet and triplet states (in the absence of the hyperfine coupling), and as a result as the applied field is increased the energy gap between the $\sing$ and $\trip_+$ states goes through zero, which leads to a maximum in the $\trip_+$ quantum yield at $\omega\simeq 2J$. However, it is not immediately obvious if and how the presence of an exchange coupling alters the low magnetic field effect on the $\trip_0$ quantum yield. In order to investigate this, we examine the $\sing$, $\trip_0$, $\trip_+$ and $\trip_-$ quantum yields for a one proton radical pair with a non-zero exchange coupling, $2J$, between the electron spins. The Hamiltonian for this radical pair is 
\begin{align}
H = \omega\left(\hat{S}_{1z}+\hat{S}_{2z}\right) + a \hat{\vb{S}}_1\cdot\hat{\vb{I}} - 2J \hat{\vb{S}}_1\cdot\hat{\vb{S}}_2.
\end{align}
Other than the inclusion of the exchange coupling, we do not alter any of the parameters used in this model, i.e. we use Eq. (17) to calculate the quantum yields with $k=a/2$. It is possible to obtain analytic expressions for the eigenstates and eigenvalues of this Hamiltonian, however the expressions are quite complicated and little is gained by analysing them. Instead we numerically calculate the quantum yields using Eqs. (3), (4) and (17) in the main text, for a range of $\omega$ and $2J$ values to investigate the effect of exchange coupling on the different triplet state quantum yields.
\begin{figure}
	\includegraphics[width=0.8\textwidth]{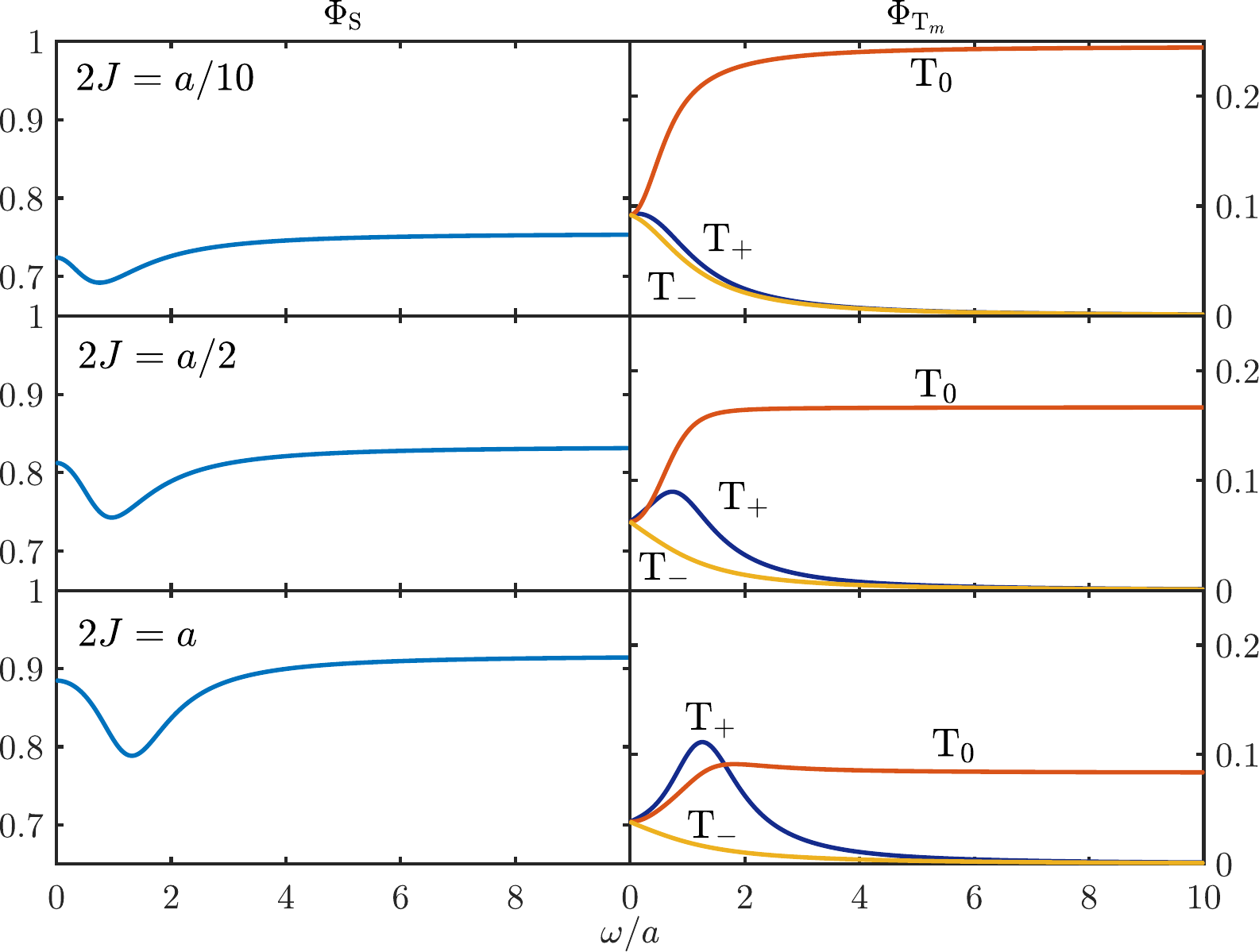}
	\caption{Singlet (left) and triplet (right) quantum yields for the one proton radical pair model with $k=a/2$, as a function of $\omega/a$, for three different values of $2J$. From top to bottom $2J=a/10,\ a/2$ and $a$. Medium blue lines are $\Phi_\sing$, red lines are $\Phi_{\trip_0}$, dark blue are $\Phi_{\trip_+}$ and gold are $\Phi_{\trip_-}$.}\label{fig1}
\end{figure}

Fig. \ref{fig1} shows singlet and triplet quantum yields for the model radical pair with $2J=a/10,\ a/2$ and $a$. As $2J$ is increased, the triplet quantum yields decrease and the singlet quantum yields increase, as energy conservation restricts the amount of S--T interconversion. Examining the individual triplet quantum yields, we see that for all values of $2J$ the $\trip_0$ quantum yield increases sharply when the field is applied before plateauing, just as in the $2J=0$ case (see Fig. 1 in the main text), i.e. the symmetry breaking still leads to an increase in $\trip_0$ production. The $\trip_+$ quantum yield goes through a maximum at $2J\simeq \omega$ and then decreases at higher fields, as the energy gap between this state and the singlet state first decreases, then increases. We also see that application of the field decreases the $\trip_-$ quantum yield, as the energy separation between the $\sing$ and $\trip_-$ states increases. The size of the maximum in the $\trip_+$ yield relative to the increase in the $\trip_0$ yield due to the low field effect increases as $2J$ increases. This is because the zero-field S--T energy gap increases with $2J$, so the effect of the crossing of the $\sing$ and $\trip_+$ states becomes larger, relative to the low magnetic field effect. 

Overall the singlet yield has the same qualitative behaviour with $2J\neq 0$ as $2J=0$: the quantum yield first decreases then increases, and plateaus at a value higher than its zero-field value when the applied field is very large. However, the minimum in $\Phi_\sing$ is now a result of both the low field effect, enhancing $\sing\to\trip_0$ transitions, and the crossing of the $\sing$ and $\trip_+$ states enhancing the $\sing\to\trip_+$ transitions. As $2J$ increases the latter becomes more important, so the minimum shifts to $\omega\simeq 2J$.